\documentclass[review,secnumarabic,preprintnumbers,amsmath,amssymb,floatfix]{revtex4}
\usepackage{graphicx}
\usepackage{subfigure} 
\usepackage{amssymb}

\begin{document}

\title{Time-Delayed Feedback in Neurosystems}

\author {E. Sch\"oll}
\author {G. Hiller}
\author {P. H{\"o}vel}
\author {M. A. Dahlem}

\affiliation{Institut f{\"u}r Theoretische Physik, Technische Universit{\"a}t Berlin,  Hardenbergstr. 36, D-10623 Berlin, Germany}

\begin{abstract}
The influence of time delay in systems of two coupled excitable neurons 
is studied in the framework of the FitzHugh-Nagumo model. Time-delay 
can occur in the coupling between neurons or in a self-feedback loop.
The stochastic synchronization of instantaneously coupled neurons under
the influence of white noise can be deliberately controlled by local
time-delayed feedback. By appropriate choice of the delay time
synchronization can be either enhanced or suppressed. In delay-coupled
neurons, antiphase oscillations can be induced for sufficiently large
delay and coupling strength. The additional application of time-delayed
self-feedback leads to complex scenarios of synchronized in-phase or 
antiphase oscillations, bursting patterns, or amplitude death. 

\end{abstract}

\maketitle

\section{Introduction}

Control of unstable or irregular states of nonlinear dynamic systems 
is a central issue of current research \cite{SCH07}. 
A particularly simple and efficient control scheme is time-delayed  
feedback which occurs naturally in a number of biological systems
including neural networks where both propagation delays of 
electrical signals connecting different neurons and 
local neurovascular couplings 
lead to time delays \cite{HAK06,WIL99,GER02}. 
Moreover, time-delayed feedback loops might be deliberately implemented to
control neural disturbances, e.g., to suppress undesired synchrony 
of firing neurons in Parkinson's disease or epilepsy \cite{SCH94e,ROS04a,POP05}.
Chaos control techniques have been first applied experimentally {\it
in vitro} in a spontaneously bursting neural network \cite{SCH94e}
using a method proposed by Ott, Grebogi and Yorke \cite{OTT90}. In
contrast to this method, the time-delayed feedback method by Pyragas
\cite{PYR92} and its extensions \cite{SOC94} do not require
detailed information on the system to be controlled, and they are 
non-invasive,
i.\,e., once the target state is reached, the control force vanishes.
However, one needs to tune the feedback gain and the delay time. Therefore,
theoretical investigations help to understand optimal choices of these
parameters.  Various global delayed feedback schemes have been
proposed as effective and robust therapy of neurological diseases with
pathological synchronization causing tremor \cite{ROS04a,POP05}.  They
have been contrasted with {\it local} delayed feedback methods
\cite{GAS07b,GAS08,DAH08}.

\begin{figure}[b!]
 \includegraphics[width=0.5\textwidth]{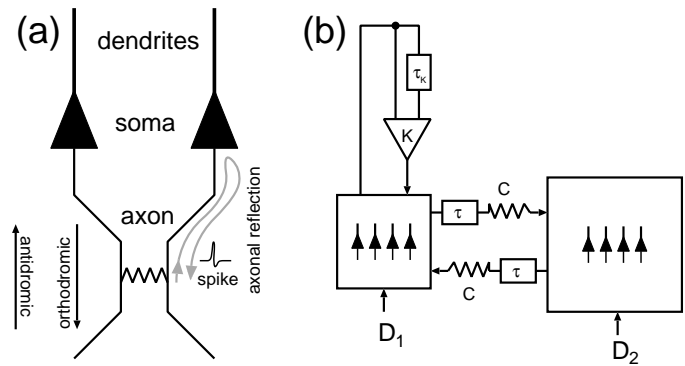} 
 \caption{(a) Scheme of two axo-axonally coupled neurons
   (pyramidal cells coupled by an electrical
   synapse) \cite{SCH01e}. (b) Two mutually coupled neural populations 
(delay $\tau$, coupling constant $C$)
   with feedback control loop (delay $\tau_K$, coupling constant $K$)
and noise input $D_1$, $D_2$.}\label{fig:scheme}
 \end{figure}

Neurons are excitable units which can emit spikes or bursts of
electrical signals, i.e., the system rests in a stable steady state,
but after it is excited beyond a threshold, it emits a pulse.  In the
following, we consider electrically coupled neurons. Such electrical
synapses are less common than chemical synapses, but it has been shown
in hippocampal slices that high-frequency synchronized oscillations
are independent of chemical synaptic transmission \cite{JEF82}, and
physiological, pharmacological, and structural evidence was provided
\cite{SCH01e} that axons of hippocampal pyramidal cells are
electrically coupled (Fig.~\ref{fig:scheme} (a)). Time delays in the 
coupling must be considered particularly in the
case of high-frequency oscillations.

The simplest model that displays features of neural interaction
consists of two coupled neural systems.  A single electrical synapse
can lead to synchronized co-operative behaviour between two
axo-axonally coupled hippocampal pyramidal cells when only one of them
is stimulated antidromically at high frequency
\cite{TRA94,TRA99}. Moreover, such neurons have been shown to reflect
antidromic spikes (propagating opposite to the normal direction) at
the soma, but a continuous reverberating activity is not generated by
axonal reflections in two axo-axonally coupled cells alone
\cite{SCH01e}.  In order to describe the complicated interaction
between billions of neurons in large neural networks, the neurons are
often lumped into highly connected sub-networks or synchronized
sub-ensembles. Such neural populations are usually localized spatially
and contain both excitatory and inhibitory neurons \cite{WIL72}. In
this sense, the model of two mutually coupled neurons may also serve
as a paradigm of two coupled neural sub-ensembles.

Starting from such simplest network motifs, larger networks can be
built, and their effects may be studied. For example, starting from
two interconnected reticular thalamic neurons with oscillatory
behavior, it was shown in \cite{DES94} how more complex dynamics
emerges in ring networks with nearest neighbors and fully reciprocal
connectivity, or in networks organized in a two-dimensional array with
proximal connectivity and ``dense proximal'' coupling in which every
neuron connects to all other neurons within some radius.  In another
example \cite{ZHO06c}, a neural population was itself modeled as a
small sub-network of excitable elements, to study hierarchically
clustered organization of excitable elements in a {\em network of
networks}.

In the following, we consider two mutually coupled neurons modelled by the
paradigmatic FitzHugh-Nagumo system \cite{FIT61,NAG62,LIN04} in the excitable regime:
  \begin{eqnarray}\label{eq:system}
    \epsilon_1 \dot{x}_1 &=& x_1 - \frac{x_1^3}{3} - y_1+ C[x_2(t-\tau)-x_1(t)]\nonumber\\
        \dot{y}_1 &=& x_1 + a + D_1 \xi_1(t) \nonumber\\
        \epsilon_2 \dot{x}_2 &=& x_2 - \frac{x_2^3}{3} - y_2 + C[x_1(t-\tau)-x_2(t)]\nonumber\\
        \dot{y}_2 &=& x_2 + a + D_2 \xi_2(t)
    \end{eqnarray}
where $x_1, y_1$ and $x_2, y_2$ correspond to single neurons (or neuron populations), which 
are linearly coupled with coupling strength $C$. The variables $x_1$, $x_2$ are related 
to the transmembrane voltage and $y_1$, $y_2$ refer to various quantities connected to 
the electrical conductance of the relevant ion currents. 
Here $a$ is an excitability parameter whose value defines whether the system is
excitable ($a>1$) or exhibits self-sustained periodic firing ($a<1$), 
$\epsilon_{1}$ and 
$\epsilon_2$ are the timescale parameters that are usually chosen to
be much smaller than unity, corresponding to fast activator variables
$x_1$, $x_2$, and slow inhibitor variables $y_1$, $y_2$. 

The synaptic coupling between two neurons is modelled
as a diffusive coupling considered for simplicity to be symmetric
\cite{LIL94,PIN00,DEM01}.  
More general delayed couplings are
considered in \cite{BUR03}. 
The coupling strength $C$ summarizes how
information is distributed between neurons.  The mutual delay $\tau$
in the coupling is motivated by the propagation delay of action
potentials between the two neurons $x_1$ and $x_2$.

Each neuron is driven by Gaussian white noise $\xi_i(t)$ $(i=1,2)$ with zero mean 
and unity variance. The noise intensities are denoted by parameters $D_1$ and $D_2$, 
respectively.

Besides the delayed coupling we will also consider delayed
self-feedback in the form suggested by Pyragas \cite{PYR92}, where the
difference $s(t)-s(t-\tau_K)$ of a system variable $s$ (e.g.,
activator or inhibitor) at time $t$ and at a delayed time $t-\tau_K$,
multiplied by some control amplitude $K$, is coupled back into the
same system (Fig.~\ref{fig:scheme}(b)).  Such feedback loops might
arise naturally in neural systems, e.g., due to neurovascular
couplings that has a characteristic latency, or due to finite
propagation speed along cyclic connections within a neuron
sub-population, or they could be realized by external feedback loops
as part of a therapeutical measure, as proposed in
Ref. \cite{POP05}. This feedback scheme is simple to implement, quite
robust, and has already been applied successfully in a real experiment
with time-delayed neurofeedback from real-time magnetoencephalography
(MEG) signals to humans via visual stimulation in order to suppress
the alpha rhythm, which is observed due to strongly synchronized
neural populations in the visual cortex in the brain \cite{HAD06}.
One distinct advantage of this method is its non-invasiveness, i.e.,
in the ideal deterministic limit the control force vanishes on the
target orbit, which may be a steady state or a periodic oscillation of
period $\tau$. In case of noisy dynamics the control force, of course,
does not vanish but still remains small, compared to other common
control techniques using external periodic signals, for instance, in
deep-brain stimulation to suppress neural synchrony in Parkinson's
disease \cite{TAS02}.

\begin{figure}[b!]
 \includegraphics[width=\textwidth]{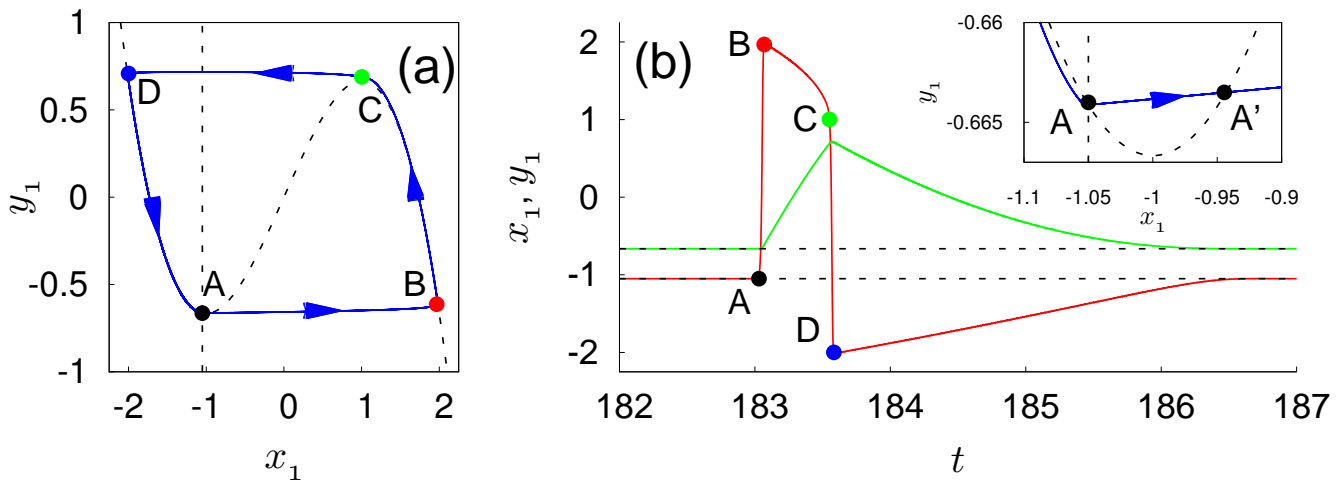} 
 \caption{\label{fig:single_FHN} 
Excitable dynamics of a single FitzHugh-Nagumo system: (a) Phase portrait $(x_1,y_1)$
(trajectory: solid blue, nullclines: dashed black),
(b) Time series of activator $x_1(t)$ (red) and inhibitor $y_1(t)$ (green). 
The colored dots A, B, C, D mark corresponding points on panels (a) and (b). 
The inset in (b) shows a blow-up of the phase portrait near A.
Parameters: $\epsilon_1=0.01$, $a=1.05$, $D_1=0$.}
 \end{figure}

The phase portrait and the null-isoclines of a single FitzHugh-Nagumo system without noise
and feedback are shown in Fig.~\ref{fig:single_FHN}(a). The fixed point A is a stable focus 
or node for $a>1$ 
({\em excitable regime}). If the system is perturbed well beyond point A' (see inset), it 
performs a large excursion $A \to B \to C \to D \to A$ in phase space corresponding to the 
emission of a spike (Fig.~\ref{fig:single_FHN}(b)). At $a=1$ the system exhibits a Hopf 
bifurcation of a limit cycle, 
and the fixed point A becomes an unstable focus for $a<1$ ({\em oscillatory regime}). 

In the following we choose the excitability parameter $a=1.05$ in the excitable regime 
close to threshold. If noise is present, it will occasionally kick the system beyond $A'$
resulting in noise-induced oscillations ({\em spiking}).

\section{Stochastic synchronization of instantaneously coupled neurons}
We shall first consider two coupled FitzHugh-Nagumo systems as in eq. (\ref{eq:system}) 
albeit without delay in the coupling ($\tau=0$). Noise can induce oscillations even though 
the fixed point is stable \cite{HAU06, HOE07}. The noise sources then play the role of 
stimulating the excitable subsystems. Even if only one subsystem is driven by noise, 
it induces oscillations of the whole system through the coupling. 
In the following, we consider two nonidentical neurons, described by different timescales
$\epsilon_1=0.005$, $\epsilon_2=0.1$, 
and set the noise intensity $D_2$ in the second subsystem equal to a small value, $D_2=0.09$, 
in order to model some background noise level. Depending on the coupling strength $C$ and the noise intensity 
$D_1$ in the first subsystem, the two neurons show cooperative dynamics. 

\begin{figure}[t]
  \includegraphics[width=0.5\linewidth]{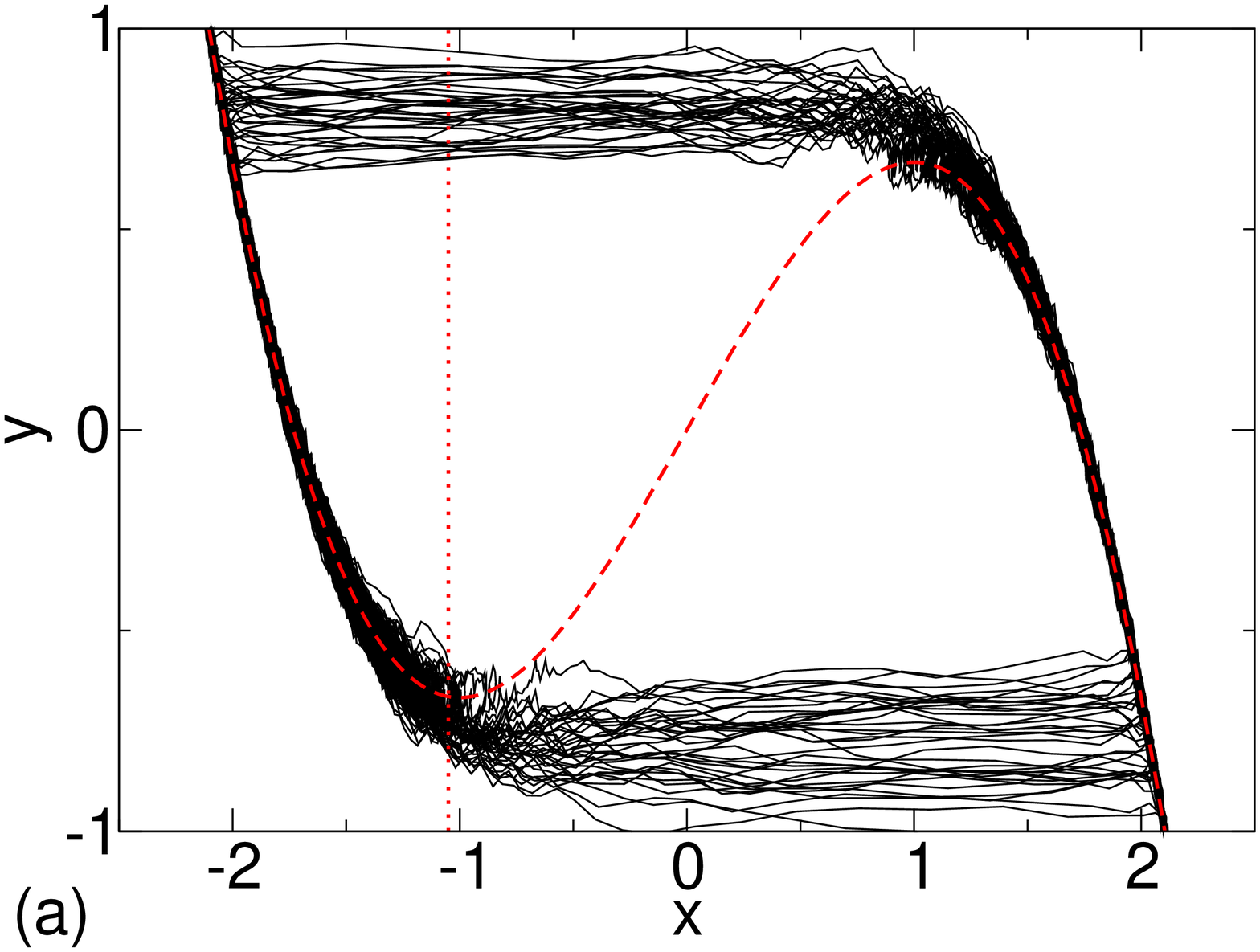}
  \includegraphics[width=0.44\linewidth]{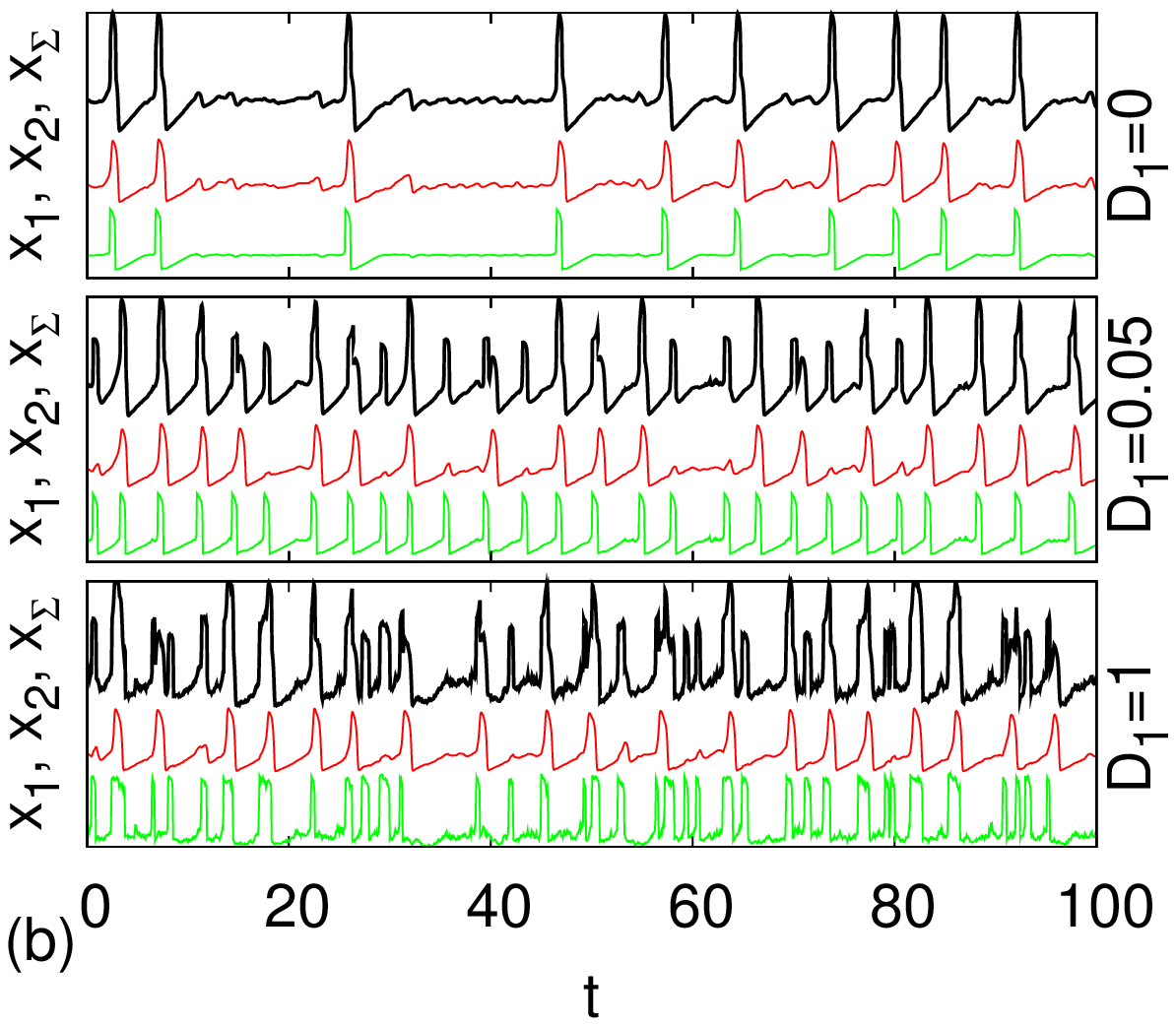}
  \caption{\label{fig1}(a): Noise-induced oscillations in a single FitzHugh-Nagumo system ($C=0$) 
for a noise intensity $D_1=0.6$.  The red dashed and dash-dotted lines denote
the null-isoclines. Panel (b): Time series of two instantaneously coupled FitzHugh-Nagumo systems 
for different noise intensities $D_1$. The  green (light grey), red (dark grey),
and black curves correspond to $x_1$, $x_2$, and their sum $x_\Sigma=x_1+x_2$, respectively. Other parameters: 
$a=1.05$, $\epsilon_1=0.005$, $\epsilon_2=0.1$, $\tau=0$, $C=0.07$ and $D_2=0.09$.}
\end{figure}

Figure~\ref{fig1}(a) depicts the temporal dynamics of a single FitzHugh-Nagumo system ($C=0)$ due to stochastic input.
One can see that the system is excitable since it performs large excursions in phase space. Figure~\ref{fig1}(b) shows
the temporal dynamics of the two instantaneously coupled neural systems for increasing noise intensities $D_1$, 
where the green (light grey), red (dark grey), and black curves correspond to $x_1$, $x_2$, and their sum
$x_\Sigma=x_1+x_2$, respectively. For $D_1=0$, the first subsystem is enslaved and emits a spike every time the second unit 
does. For increasing $D_1$, however, this synchronization is weakened and for large noise intensities $D_1$ the dynamics 
of the first subsystem is independently dominated by its own stochastic input.

\begin{figure}
  \includegraphics[width=0.4\linewidth]{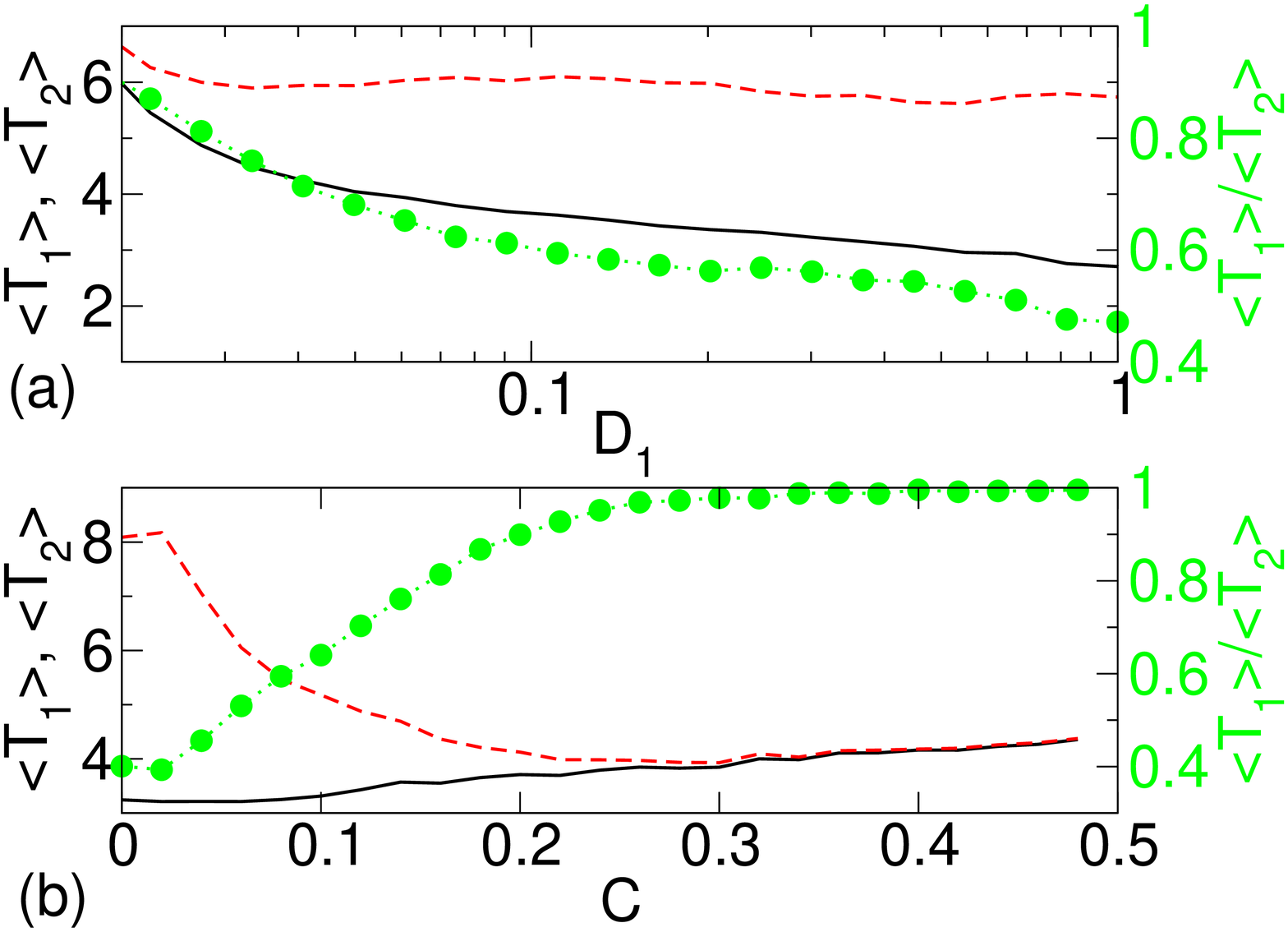}
  \includegraphics[width=0.5\linewidth]{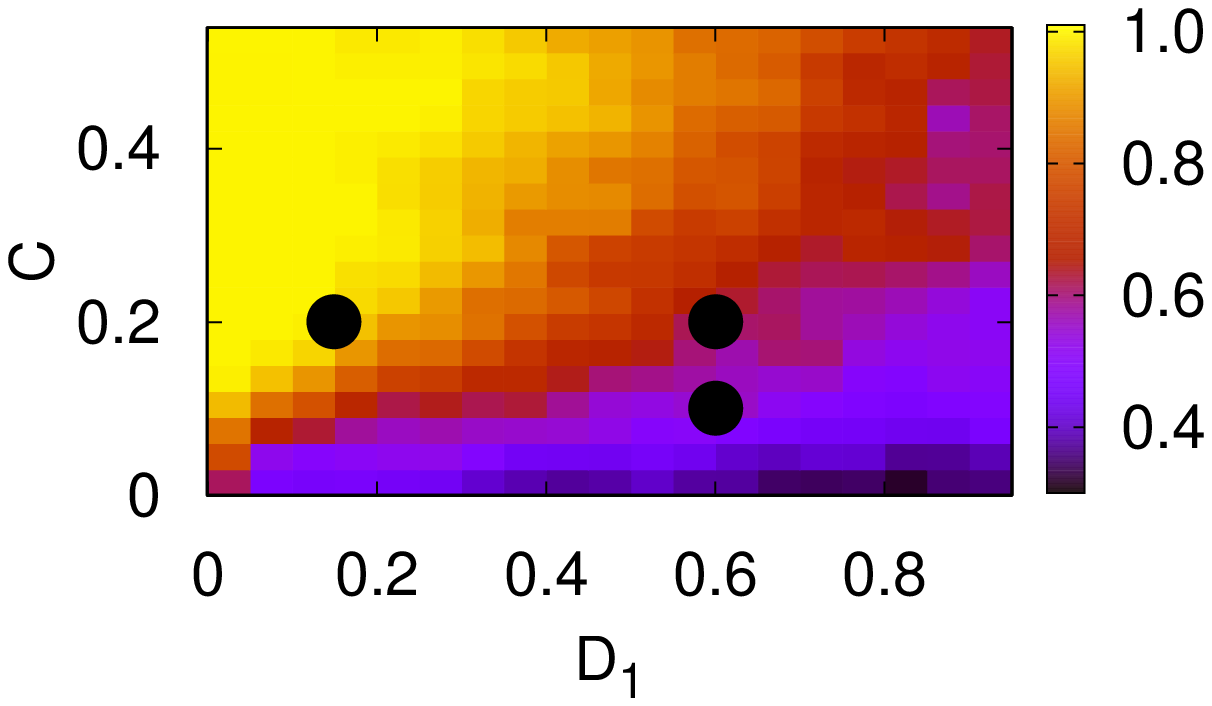}
  \caption{\label{fig2} Frequency synchronization: Average interspike intervals of each subsystem
(black solid curve: $\langle T_1\rangle$, red dashed: $\langle T_2\rangle$) and their ratio (green dotted: $\langle
T_1\rangle /\langle T_2\rangle $) (a) vs. noise intensity $D_1$ (for fixed $C=0.07$),
(b) vs. coupling strength $C$ (for fixed $D_1=0.25$). (c): Ratio of the average
interspike intervals in the ($D_1,C$) plane. The black dots mark the parameter choices for different synchronization 
regimes used in Fig.~\ref{fig:isi_tau}. Other parameters: 
$a=1.05$, $\epsilon_1=0.005$, $\epsilon_2=0.1$, $\tau=0$, and $D_2=0.09$.}
\end{figure}

There are various measures of the synchronization of coupled systems \cite{ROS01a}. 
For instance, one can consider the average
interspike intervals of each subsystem, \textit{i.e.} $\langle T_1\rangle $ and $\langle T_2\rangle $, calculated from
the $x$ variable of the respective subsystem.  Their ratio $\langle T_1\rangle /\langle T_2\rangle $ is a measure of
frequency synchronization, as depicted in
Fig.~\ref{fig2}. Panel (a) displays the ratio of average interspike intervals in dependence on the noise
intensity $D_1$ (green dots) for fixed coupling strength $C=0.07$. One can see that for increasing $D_1$ the ratio
$\langle T_1\rangle /\langle T_2\rangle $ decreases. Thus, the two subsystems become less synchronized. Panel (b) 
shows the dependence on the coupling strength $C$ for fixed noise intensity $D_1=0.25$. Without coupling ($C=0$), the
two subsystems operate on their own timescale. For increasing $C$, however, they synchronize as the ratio of
the average interspike intervals approaches unity. Panel (c) shows the
result as a function of $D_1$ and $C$, where the bright regions indicate a strongly
synchronized behaviour of the two subsystems. For small $D_1$ and large coupling strength $C$, the two subsystems
display synchronized behaviour, $\langle T_1\rangle/\langle T_2\rangle\approx 1 $. On average, they show the same number
of spikes indicated by bright (yellow) color. In the next Section, we consider three different regimes of synchronization:
moderately, weakly, and strongly synchronized systems as marked by black dots in Fig.~\ref{fig2}(c).\\

Other measures for stochastic synchronization are given by the phase synchronization index \cite{HAU06}, or the mean 
phase synchronization intervals \cite{HOE08}, but they exhibit qualitatively similar behaviour.

\section{Control of synchronization by time-delayed feedback}

In this Section, we consider the control of global cooperative dynamics by local application of a stimulus to a single
system. This stimulus is realized by time-delayed feedback, which was initially introduced by Pyragas in order to
stabilize periodic orbits in deterministic systems \cite{PYR92}:
\begin{subequations}
\begin{eqnarray}
	\label{eq:2FHN_1}
	\epsilon_1 \dot{x}_1 &=& x_1 -\frac{x_1^3}{3} -y_1 + C [x_2(t) -x_1(t)]\\
	\dot{y}_1 &=& x_1 + a + K \left[y_1(t-\tau_K) - y_1(t) \right] + D_1 \xi_1(t) \nonumber\\
	\label{eq:2FHN_2}
	\epsilon_2 \dot{x}_2 &=& x_2 -\frac{x_2^3}{3} -y_2 + C [x_1(t) -x_2(t)]\\
	\dot{y}_2 &=& x_2 + a + D_2 \xi_2(t). \nonumber	
\end{eqnarray}
\end{subequations}
The parameters of the time-delayed feedback scheme are the feedback gain $K$ and the time delay $\tau_K$. With this 
method, a control force is constructed from the differences of the states of the system which are one time unit $\tau_K$
apart. One could also consider application of the feedback scheme to both subsystems and effects of
different values of the control parameters for each subsystem, but these investigations are beyond of the scope of this
work. Previously, time-delayed feedback has also been used to influence noise-induced oscillations of a single
excitable system \cite{JAN03,BAL04,PRA07}, of systems below a Hopf bifurcation \cite{SCH04b,POM05a,POT07,FLU07} or 
below a global bifurcation \cite{HIZ06,HIZ08}, and of spatially extended reaction-diffusion systems 
\cite{STE05a,BAL06,DAH08}. Extensions to multiple time-delay control schemes have also been considered 
\cite{POM07,SCH08a,HOE07,HOE08}.

\begin{figure}[th]
  \begin{center}
	\includegraphics[width=0.32\linewidth]{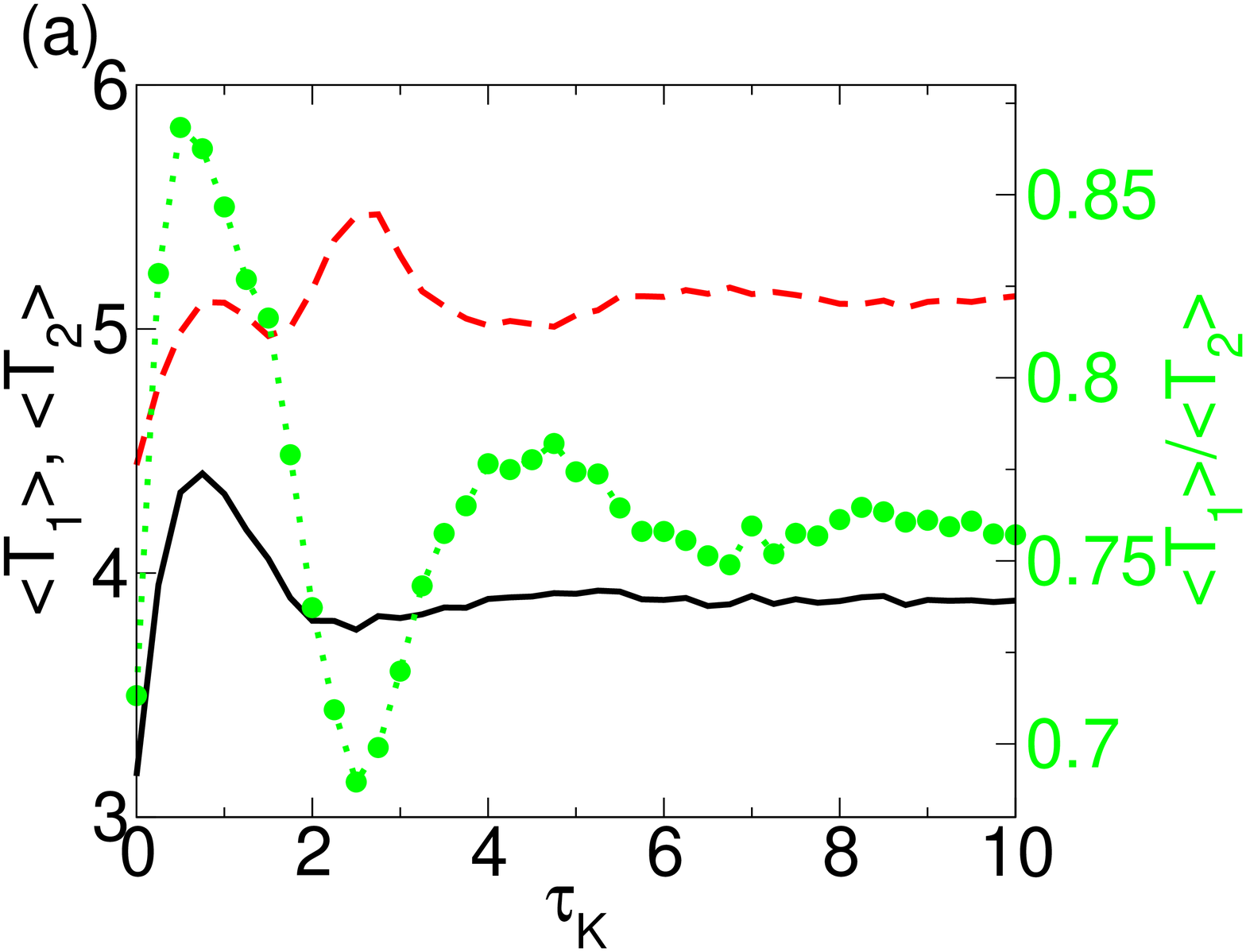}
	\includegraphics[width=0.32\linewidth]{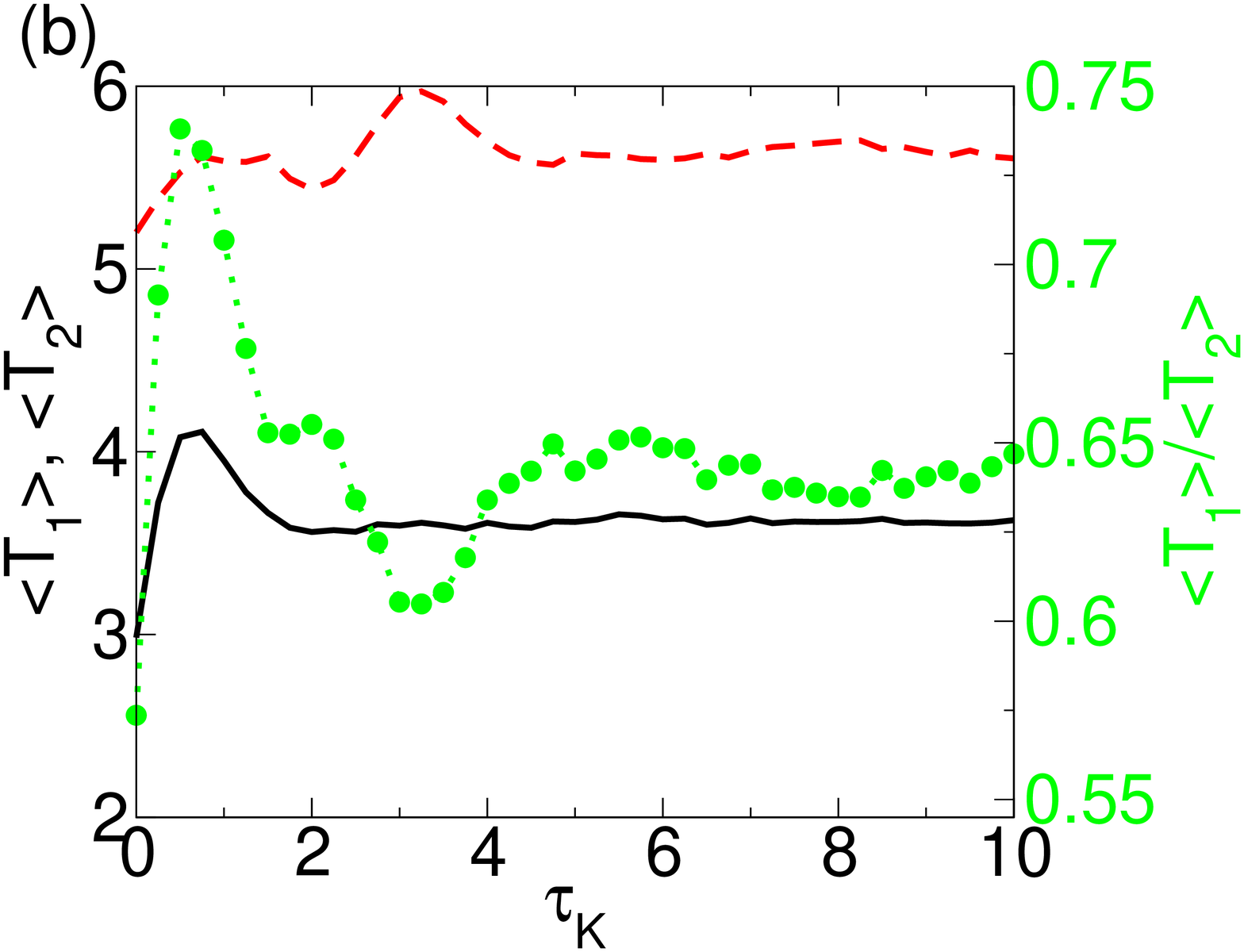}
	\includegraphics[width=0.32\linewidth]{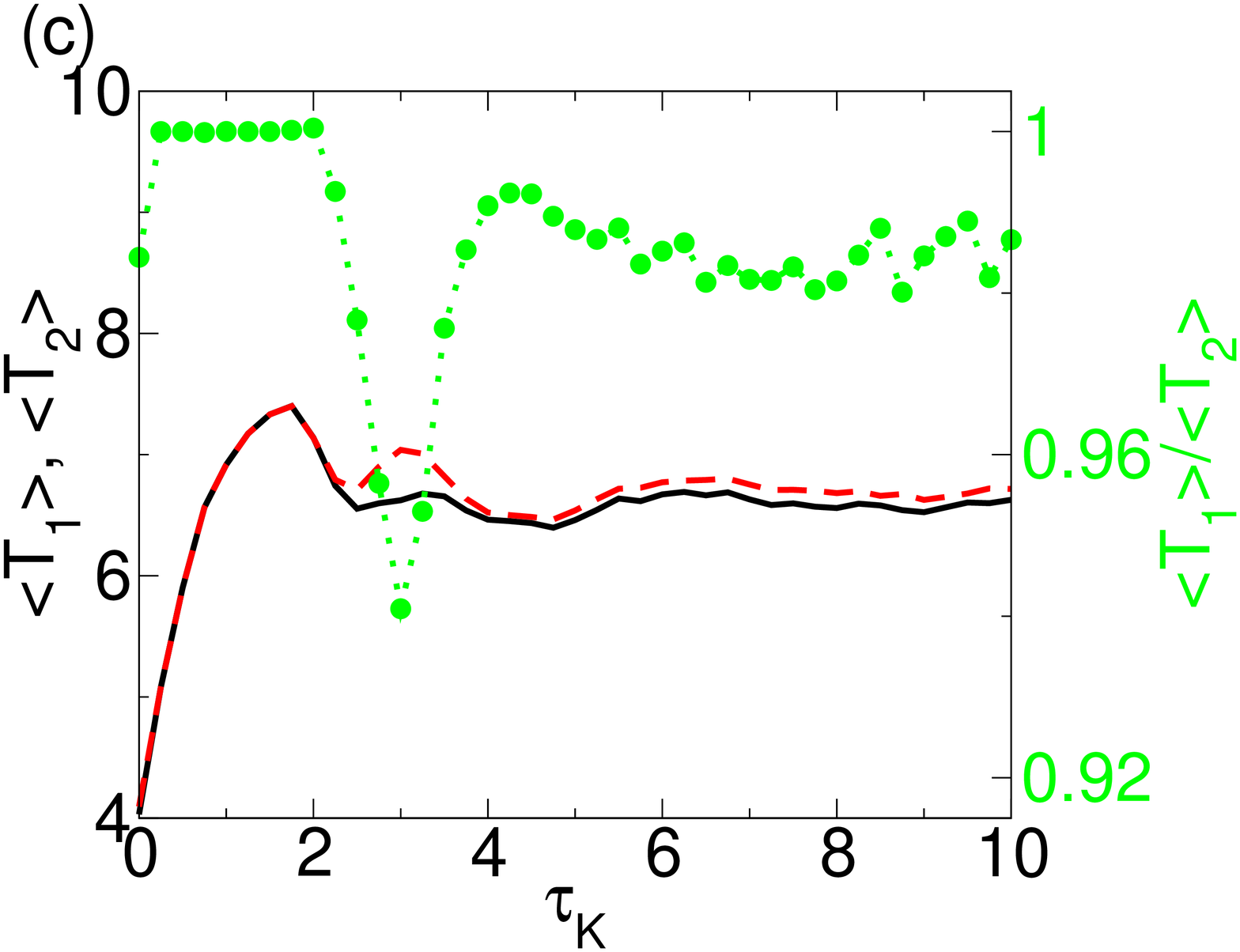}
  \end{center}
  \caption{\label{fig:isi_tau} Control of synchronization: Interspike intervals vs. time delay 
$\tau_K$. Green dots: ratio of the interspike intervals $\langle T_1\rangle /\langle T_2\rangle $ of the two
subsystems, black solid curve: $\langle T_1\rangle$, red dashed curve: $\langle T_2\rangle $. 
Panels (a), (b), and (c) correspond to the case of moderate ($C=0.2$, $D_1=0.6$), weak ($C=0.1$, $D_1=0.6$), 
and strong ($C=0.2$, $D_1=0.15$) synchronization, respectively. Other parameters:
$a=1.05$, $\epsilon_1=0.005$, $\epsilon_2=0.1$, $C=0.07$, $\tau=0$, $D_2=0.09$, $K=1.5$.}
\end{figure}

\begin{figure}[th]
  \begin{center}
	\includegraphics[width=\linewidth]{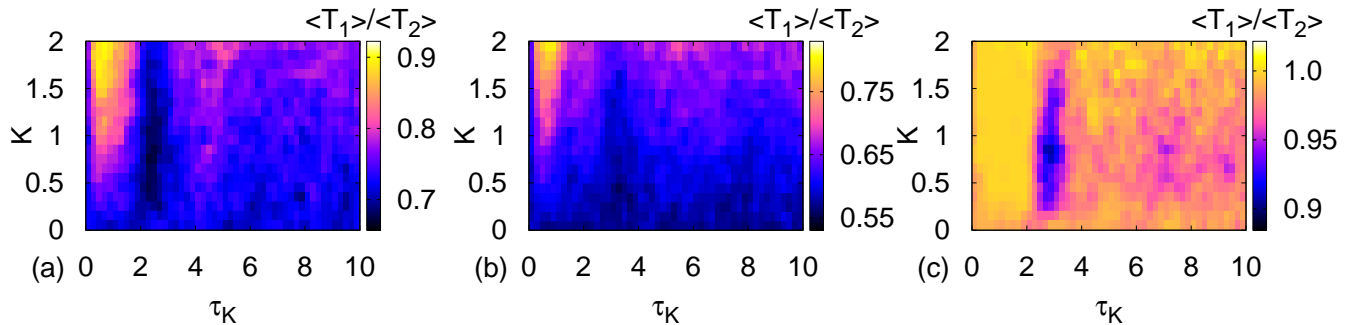}
  \end{center}
  \caption{\label{fig:isi_control} Ratio of interspike intervals $\langle T_1\rangle / \langle
T_2\rangle$ in dependence on the feedback gain $K$ and the time delay $\tau$ for (a) moderate ($C=0.2$, $D_1=0.6$), 
(b) weak ($C=0.1$, $D_1=0.6$), and (c) strong ($C=0.2$, $D_1=0.15$) synchronization. 
Other parameters as in Fig.~\ref{fig2}(c).}
\end{figure}
Since we are interested in the effects of a control force on the synchronization, in the following we consider three
different cases: moderately, weakly, and strongly synchronized systems given by the specific choices of the coupling
strength and noise intensity in the first subsystem.
These different cases of stochastic synchronization are marked as black dots in Fig.~\ref{fig2}. 

As a measure to quantify changes in the synchronization due to the control force, we consider the ratio of average
interspike intervals. In the presence of a control force, i.e., $K\neq 0$, the cooperativity can be influenced by 
varying the feedback gain $K$ and the time delay $\tau_K$.

For fixed feedback gain $K=1.5$, Fig.~\ref{fig:isi_tau} depicts the average interspike intervals of the subsystems,
shown as solid (black) and dashed (red) curves for $\langle T_1\rangle$ and $\langle T_2\rangle $, and their ratio 
(green dots) for the case of (a) moderately, (b) weakly, and (c) strongly synchronized systems, respectively, in dependence
on the time delay $\tau_K$. In all three cases, the stochastic synchronization can be strongly modulated by changing the 
delay time, i.e., one can either enhance and suppress synchronization by appropriate choice of the local feedback delay.

The overall dependence of the frequency synchronization, measured by the ratio of 
$\langle T_1\rangle/\langle T_2\rangle$, is displayed in dependence on the control parameters $K$ and $\tau_K$
in Fig.~\ref{fig:isi_control} for moderate, weak, strong synchronization in panels (a), (b), and
(c), respectively. Thus, Fig.~\ref{fig:isi_tau} can be understood as a horizontal cut for $K=1.5$ through
Fig.~\ref{fig:isi_control}. One can see a modulation of the ratio of average interspike interval by $\tau_K$ for a
large range of feedback gain. In view of applications, where neural synchronization is often pathological, 
e.g. in Parkinson's disease or epilepsy, it is interesting to note that there are cases where a proper choice 
of the local feedback control parameters leads to desynchronization of the coupled system 
(dark regions in Fig.~\ref{fig:isi_control}).

\section{Delay-coupled neurons}

In this section we study the influence of a delay in the coupling of two neurons, rather than a
delayed self-feedback.
We set the noise terms in Eq.~(\ref{eq:system}) equal to zero, $D_1=D_2=0$, 
but consider a time-delay $\tau$ in the 
coupling. In the deterministic system the delayed coupling plays the role of a stimulus
which can induce self-sustained oscillations in the coupled system even if the fixed
point is stable. In this sense the delayed coupling has a similar effect as the noise 
term in the previous sections. Here the bifurcation parameters for delay-induced
bifurcations are the coupling parameters $C$ and $\tau$.  

\subsection{Linear stability of fixed point}

In the following we shall choose symmetric timescales $\epsilon_1=\epsilon_2=\epsilon=0.01$ 
and fix $a=1.05$, where each of the two subsystems has a stable fixed point and exhibits excitability.

The unique fixed point of the system is symmetric and is given by 
${\bf x}^* \equiv (x_1^*, y_1^*, x_2^*, y_2^*)$, where $x_i^* = -a$, $y_i^*= a^3/3 - a$.
Linearizing Eq. (\ref{eq:system}) around the fixed point $\bf{x}^*$ by setting 
${\bf x}(t)={\bf x}^*+\delta{\bf x}(t)$, one obtains:
\begin{eqnarray}
\delta\dot{\bf{x}}
= \frac{1}{\epsilon} \left( \begin{array}{cccc}
\xi & -1 & 0 & 0\\
\epsilon & 0 & 0 & 0\\
0 & 0 & \xi & -1\\
0 & 0 & \epsilon & 0\end{array} \right)
\delta{\bf x}(t) 
+ \frac{1}{\epsilon} \left( \begin{array}{cccc}
0 & 0 & C & 0\\
0 & 0 & 0 & 0\\
C & 0 & 0 & 0\\
0 & 0 & 0 & 0\end{array} \right)	
\delta{\bf x}(t-\tau)
\end{eqnarray}
where $\xi=1-a^2-C$.
The ansatz
\begin{equation}\delta\bf{x}(t)=e^{\lambda t} \bf{u}\end{equation}
where $\bf{u}$ is an eigenvector of the Jacobian matrix, 
leads to the characteristic equation for the eigenvalues $\lambda$: 
\begin{equation}
(1- \xi \lambda+\epsilon \lambda^2 )^2-(\lambda C e^{-\lambda \tau })^2=0,
\end{equation}
which can be factorized giving 
\begin{equation}
1- \xi \lambda+\epsilon \lambda^2 \pm \lambda C e^{-\lambda \tau }=0.
\label{char_eq}
\end{equation}

\begin{figure}[b!]
\includegraphics[width=0.5\textwidth]{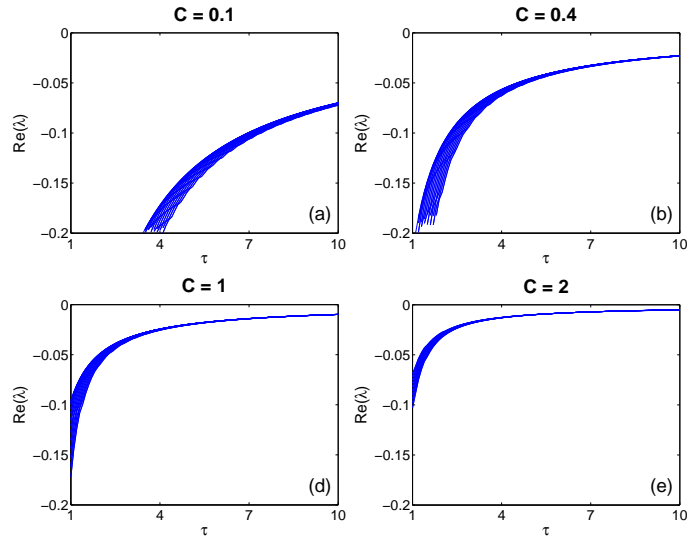} 
\caption{\label{fig:eigenmodes} Real parts Re($\lambda$) of the eigenvalues 
of the fixed point vs. time delay $\tau$ for $a\!=\!1.05$, $\epsilon=0.01$,  
and (a) C=0.1, (b) C=0.4, (c) C=1, (d) C=2.}
\end{figure}

This transcendental equation has infinitely many complex solutions $\lambda$. 
Fig.~\ref{fig:eigenmodes} shows the real parts of $\lambda$ for various values of $C$.
As can be seen in Fig.~\ref{fig:eigenmodes} 
the real parts of all eigenvalues are negative throughout, 
i.\,e., the fixed point of the coupled system remains stable for all $C$.  
This can be shown analytically for $a>1$ by demonstrating that no delay-induced
Hopf bifurcation can occur. Substituting the ansatz $\lambda=i \omega$ into 
Eq.~(\ref{char_eq}) and separating into real and imaginary parts yields for the 
imaginary part
\begin{equation}
\xi =\pm C \cos(\omega \tau)
\end{equation}
This equation has no solution for $a>1$ since $|\xi|=a^2-1+C>C$, 
which proves that a Hopf bifurcation cannot occur.

\subsection{Delay-induced antiphase oscillations}
\begin{figure}[t!]
\includegraphics[width=0.7\textwidth]{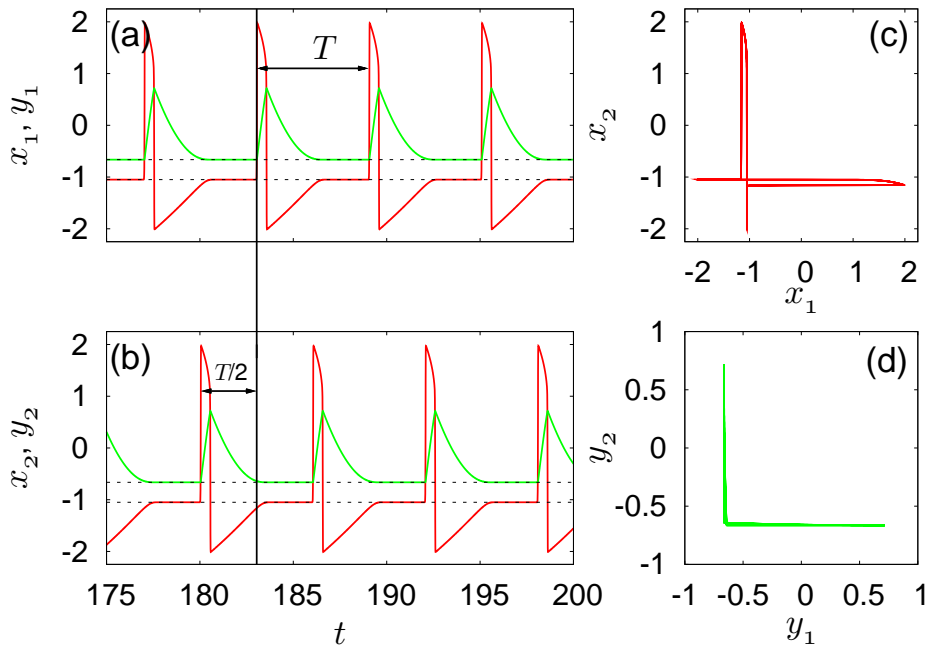} 
\caption{\label{fig:ts_comparison} Delay-induced oscillations. (a),
  (b): Time series of both subsystems (red solid lines: activator
  $x_i$, green solid lines: inhibitor $y_i$; black dashed lines: fixed
  point values of activator and inhibitor).  (c), (d):
  Phase portraits of
  activators (c) and inhibitors (d).  Parameters: $a\!=\!1.05$, $\epsilon=0.01$,
  $C=0.5$, $\tau=3$. }
\end{figure}

Delay-induced oscillations in excitable systems are inherently different from
noise-induced oscillations.  The noise term continuously kicks the subsystems out
of their respective rest states, and thus induces sustained oscillations.
Instantaneous coupling without delay 
then produces synchronization effects between the individual
oscillators \cite{HAU06,HOE07,HOE08}.  
For {\em delayed} coupling the case is entirely different.  Here the 
impulse of one neuron triggers the other neuron to emit a spike,
which in turn, after some delay, triggers the first neuron to emit 
a spike. Hence self-sustained periodic oscillations can be
induced without the presence of noise (Fig.~\ref{fig:ts_comparison}).  
It is evident that the oscillations of the two neurons have a phase lag of $\pi$.
The period of the oscillations is given by $T=2(\tau + \delta)$ with a small
quantity $\delta>0$. 

In order to understand this additional phase-shift $\delta$, we
shall now consider in detail the different stages of the oscillation as 
marked in Fig.~\ref{fig:single_FHN}. Due to the small value of $\epsilon \ll 1$
there is a distinct timescale separation between the fast activators and
the slow inhibitors, and a single FitzHugh-Nagumo system performs a 
fast horizontal transition $A \to B$, then travels slowly approximately along the
right stable branch of the $x_1$ nullcline  $B \to C$ ({\em firing}), 
then jumps back fast to $D$, and returns
slowly to the {\em rest state $A$} approximately along the left stable branch of the $x_1$ nullcline
({\em refractory phase}). 
If $a$ is close to unity, these four points are approximately given by
$A=(-a, -a+\frac{a_1^3}{3})$, $B=(2,-\frac{2}{3})$, $C=(1,\frac{2}{3})$, $D=(-2,\frac{2}{3})$.
A rough estimate for $A'$ is $(a-2,-a+\frac{a_1^3}{3})$.
The two slow phases $B \to C$ and $D \to A$ can be approximated by
$y_1 \approx x_1 - \frac{x_1^3}{3}$ and hence  $\dot{y}_1 \approx \dot{x}_1(1-x_1^2)=x_1 + a$
which gives 
\begin{eqnarray}\label{eq:slow}
   \dot{x}_1 &=&  \frac{x_1+a}{1-x_1^2}
\end{eqnarray}
which can be solved analytically, describing the firing phase (+) and the refractory phase (-):
\begin{eqnarray}\label{eq:T_fire0}
  \int_{\pm 2}^x dx_1 \frac{1-x_1^2}{x_1+a}=(a^2-1) \ln\frac{a\pm 2}{a+x}-a(\pm 2-x)+2
-\frac{x^2}{2} = t.
\end{eqnarray}
Integrating from $B$ to $C$ gives the firing time
\begin{eqnarray}\label{eq:T_fire}
T_f=\int_2^1 dx_1 \frac{1-x_1^2}{x_1+a}=(a^2-1) \ln\frac{a+2}{a+1}-a+\frac{3}{2}.
\end{eqnarray}
For $\epsilon=0.01$, $a=1.05$ the analytical solution is in good
agreement with the numerical solution in Fig.~\ref{fig:single_FHN}(b),
including the firing time $T_f=0.482$ 
(analytical approximation: $0.491$).

For a rough estimate, in the following we shall approximate the spike by a rectangular pulse
\begin{equation}\label{eq:spikeactivator}
  x_1(t) \approx \left\{ \begin{array}{ll}
  2& \mbox{if }\,\, t<T_f\\
  -a& \mbox{if }\,\, t\ge T_f
\end{array} \right.
\end{equation}

If the first subsystem is in the rest state, and a spike of the second subsystem
arrives at $t=0$ (after the propagation delay $\tau$), we can approximate the
initial dynamic response by linearizing $x_1, y_1$ around the fixed point
$(x_1^*, y_1^*)$ and approximating the feedback by a constant impulse during
the firing time $T_f$.
The fast dynamic response along the $x_1$ direction is then given by
\begin{eqnarray}\label{eq:fast}
   \epsilon \delta\dot{x}_1 &=&  \xi \delta x_1+ 2C
\end{eqnarray}
with $\xi<0$.
This inhomogeneous linear differential equation can be solved with the initial condition
$x_1(0)=-a$:
\begin{eqnarray}\label{eq:fast1}
   x_1(t) &=&  -a + \frac{2C}{|\xi|} (1-e^{-\frac{|\xi|}{\epsilon}t})
\end{eqnarray}
Note that this equation is not valid for large $t$ since (i) the linearization breaks down,
and (ii) the pulse duration $T_f$ is exceeded.
For small $t$ Eq.(\ref{eq:fast1}) can be expanded as
\begin{eqnarray}\label{eq:fast2}
   x_1(t) &=&  -a + \frac{2C}{\epsilon} t
\end{eqnarray}
which is equivalent to neglecting the upstream flow field $-|\xi| \delta x_1$ in 
Eq.(\ref{eq:fast}) near the stable fixed point $A$ compared to the pulling force $2C$ of
the remote spike which tries to excite the system towards $B$. Once the system has
crossed the middle branch of the $x_1$ nullisocline at $A'$, the intrinsic flow field 
accelerates the trajectory fast towards $B$, initiating the firing state.
Therefore there is a turn-on delay $\delta$, given by the time the trajectory takes
from $A$ to $A'$, i.e. $x_1(\delta)\approx a-2$, according to Eq.(\ref{eq:fast2}):
\begin{eqnarray}\label{eq:delta}
   \delta &=&  (a-1)\frac{\epsilon}{C}
\end{eqnarray}
Since the finite rise time of the impulse has been neglected in our estimate,
the exact solution $\delta$ is slightly larger and does not vanish at $a=1$.

\begin{figure}[b!]
\includegraphics[width=0.5\textwidth]{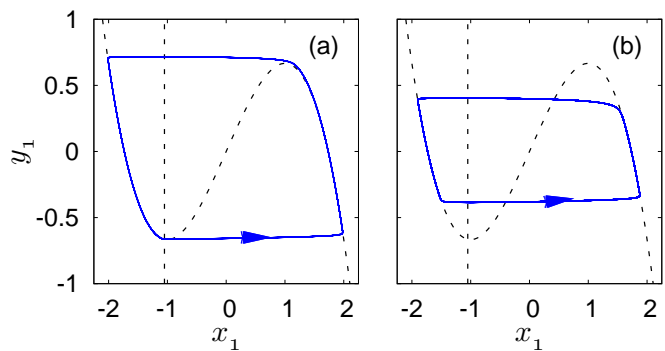} 
\caption{\label{fig:phaseportraits}  
Phase portraits of delay-coupled excitable system $(x_1,y_1)$  
for different delay times $\tau$
(trajectories: solid blue, nullclines: dashed black).
(a) $\tau \!=\! 3$ ($\delta =0.009$), 
(b) $\tau\!=\!0.8$ ($\delta =0.015$).  
Other parameters: $a\!=\!1.05$, $\epsilon\!=\!0.01$, $C\!=\!0.5$.}
\end{figure}

With increasing $a$ the distance $A-A'$ increases, and so does $\delta$.
The small additional phase shift $\delta$ between the spike $x_1(t)$ and the delayed pulse
$x_2(t-\tau)$ results in a non-vanishing coupling term at the beginning and at the end of the spike
$x_1(t)$. It is the reason (i) that the spike is initiated, 
and (ii) that it is terminated slightly before the turning point of the $x_1$ nullcline.
The latter effect becomes more pronounced if $a$ is increased or $\tau$ is decreased (Fig. 
\ref{fig:phaseportraits}). Both lead to a shift of the initial starting point of the 
spike emission on the left branch of the nullcline towards $D$, and hence to a longer 
distance up to the middle branch of the nullcline which has to be overcome by the impulse $x_2$,
hence to a larger turn-on delay $\delta$, and therefore to an earlier termination of the spike
$x_1$. This explains that the firing phase is shortened, and the limit cycle loop is 
narrowed from both sides with increasing $a$ or decreasing $\tau$, see Fig. 
\ref{fig:phaseportraits}. 
In the case of $a=1.05$ and $\tau=3$ (Fig.~\ref{fig:phaseportraits}
(a)), the delay time is large enough for the two
subsystems to nearly approach the fixed point $A$ before being perturbed
again by the remote signal.
If the delay time becomes much smaller, e.\,g., for $\tau=0.8$ (Fig.~\ref{fig:phaseportraits} (b)),
the excitatory spike of the other subsystem arrives while the first system is still
in the refractory phase, so that it cannot complete the return $D \to A$ 
to the fixed point. In this case, $a$ in Eq.(\ref{eq:delta}) has to be substituted by a 
larger value $\tilde{a}$ with $a<\tilde{a}<1.7$ in order to get a better estimate of $\delta$.
Note that without the phase-shift $\delta$ the coupling term
$C[x_2(t-\tau)-x_1(t)]$ would always vanish in the $2\tau$-periodic state.

Next, we shall investigate conditions upon the coupling parameters $C$ and $\tau$
allowing for limit cycle oscillations. On one hand, if $\tau$ becomes smaller than
some $\tau_{min}$, the impulse from the excitatory neuron arrives too early to trigger
a spike, since the system is still early in its refractory phase. On the other hand,
if $C$ becomes too small, the coupling force of the excitatory neuron is too weak to
excite the system above its threshold and pull it far enough towards $B$. 

\begin{figure}[b!]
\includegraphics[width=0.5\textwidth]{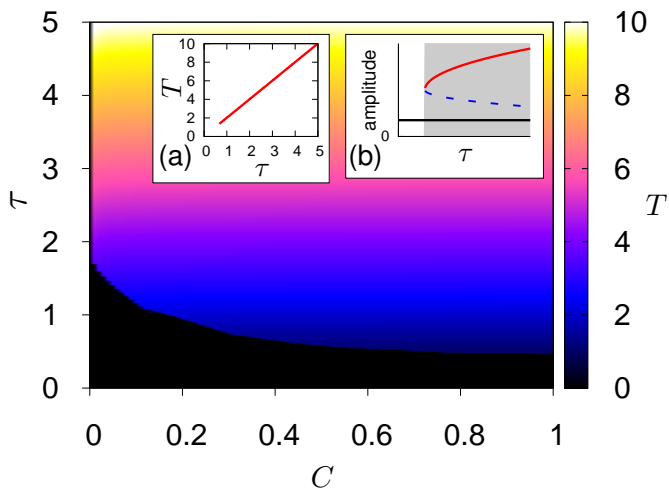} 
\caption{\label{fig:hauplot} Regime of oscillations in the $(\tau, C)$ parameter plane 
for initial conditions corresponding to single-pulse-excitation in one system. The 
oscillation period $T$ is color coded. The transition between black and color marks the 
bifurcation line.  Inset (a) shows the oscillation period vs. $\tau$ in a cut at $C=0.8$.
Inset (b): schematic plot of the saddle-node bifurcation of a stable (red solid line) and unstable
(blue dashed) limit cycle. The maximal oscillation amplitude is
plotted vs. the delay time $\tau$ and the stable fixed point is
plotted as a solid black line. The grey background marks the
bistable region.
Parameters: $a=1.05$, $\epsilon=0.01$. 
}
\end{figure} 

\begin{figure}[b!]
\includegraphics[width=.5\textwidth]{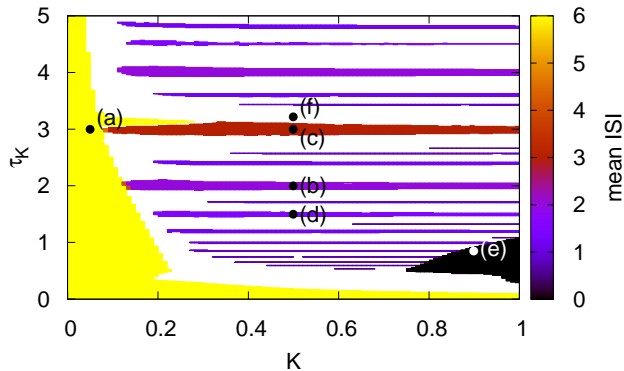} 
\caption{\label{fig:isiplane} 
Influence of delayed self-feedback upon coupled oscillations. 
The mean interspike interval (ISI) is color coded in the control parameter plane of the self-feedback
gain $K$ and delay $\tau_K$. White areas mark regimes of irregular 
oscillations where the ISI variance becomes large ($>0.01$). 
Time series corresponding to points (a)-(f) are shown in 
Fig.~\ref{fig:controlcases}.
Other parameters:  $a=1.3$, $\epsilon=0.01$, $C=0.5$, $\tau=3$.
}
\end{figure}

\begin{figure}[t!]
\includegraphics[width=.9\textwidth]{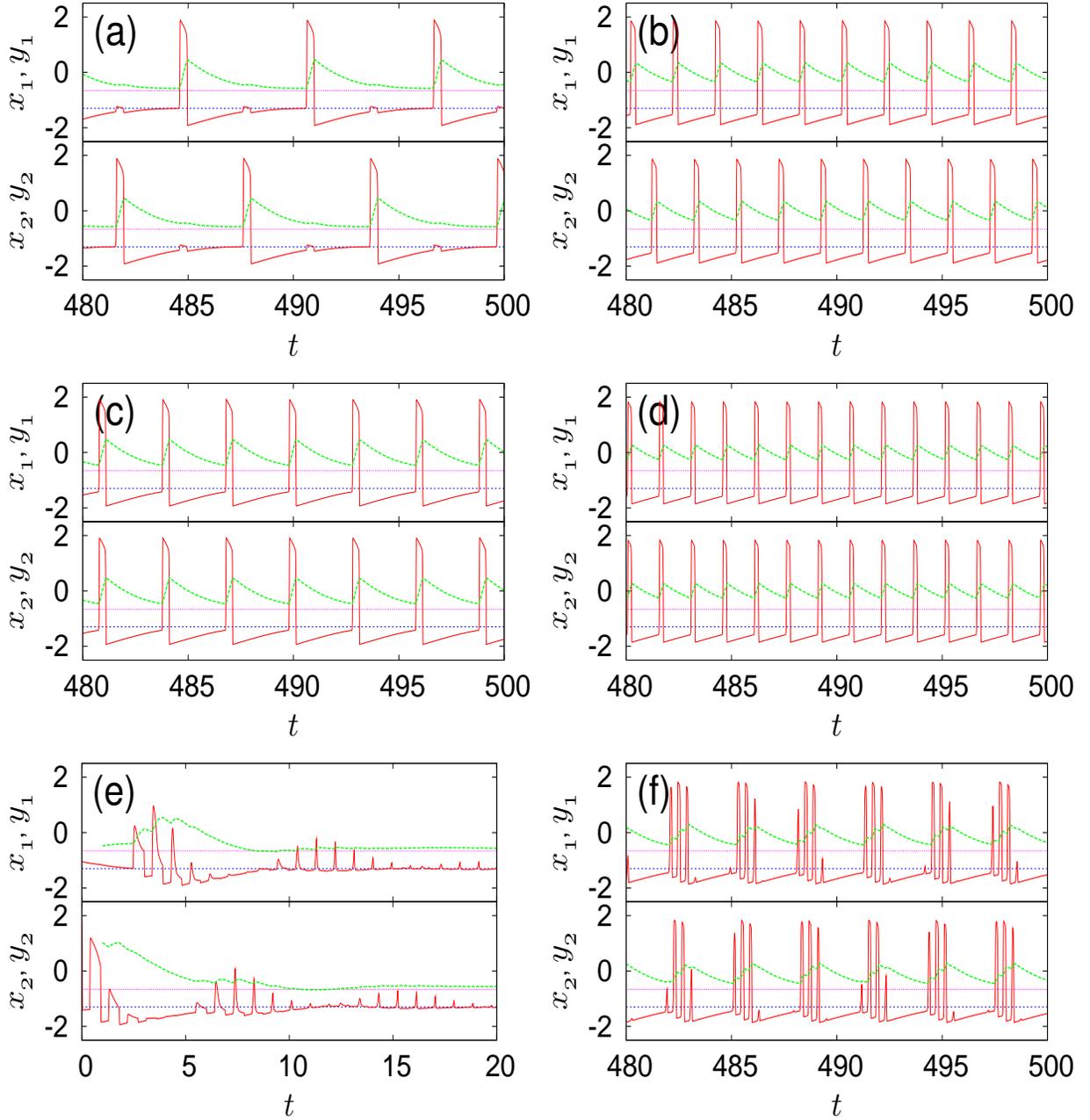} 
\caption{\label{fig:controlcases} 
Different modes of oscillation corresponding to different self-feedback parameters $K$, $\tau$
(red solid lines: activators $x_i(t)$, green solid lines: inhibitors $y_i(t)$).
(a), (b): Antiphase oscillations for (a) $K=0.05, \tau_K=3$ (period $T=6$) and 
(b) $K=0.5, \tau_K=2$ ($T=2$);  
(c), (d): In-phase oscillations for (c) $K=0.5, \tau_K=3$ (period $T=3$) and 
(d) $K=0.5, \tau_K=1.5$ ($T=1.5$);  
(e): Oscillator death for $K=0.9, \tau_K=0.9$;
(f): Bursting pattern for $K=0.5, \tau_K=3.2$. 
Other parameters:  $a=1.3$, $\epsilon=0.01$, $C=0.5, \tau=3$.
}
\end{figure}

\begin{figure}[t!]
\includegraphics[width=.5\textwidth]{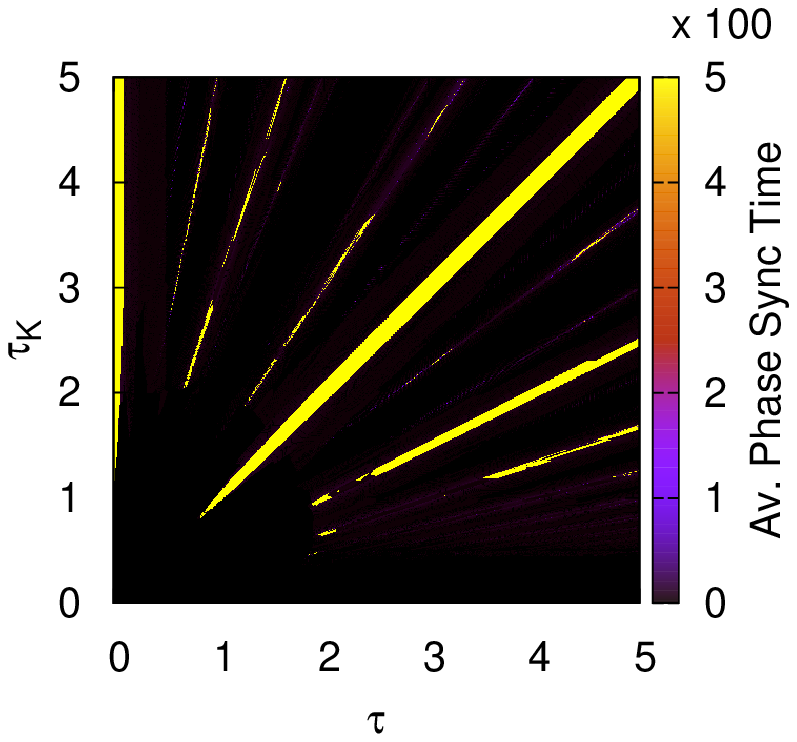} 
\caption{\label{fig:aps} 
Average phase synchronization time (color coded) in the control parameter
plane of coupling delay $\tau$ and self-feedback delay $\tau_K$.
Other parameters:  $a=1.3$, $\epsilon=0.01$, $C=0.5$, $K=0.5$.
}
\end{figure}

In Fig.~\ref{fig:hauplot} the regime of oscillations is shown
in the parameter plane of the coupling strength $C$ and coupling delay $\tau$. The
oscillation period is color coded. The boundary of this colored region is given by
the minimum coupling delay $\tau_{min}$ as a function of $C$. For large coupling strength,
$\tau_{min}$ is almost independent of $C$; with decreasing $C$ it sharply increases, and
at some small minimum $C$ no oscillations exist at all. 
At the boundary, the oscillation sets in with finite frequency and amplitude as can be 
seen in the insets of Fig.~\ref{fig:hauplot} which show a cut of the parameter plane at $C=0.8$.
The oscillation period increases linearly with $\tau$.  The mechanism that 
generates the oscillation is a saddle-node bifurcation of limit cycles 
(see inset (b) of Fig.~\ref{fig:hauplot}), creating a pair of a stable and an unstable limit cycle. 
The unstable limit cycle separates the two attractor
basins of the stable limit cycle and the stable fixed point.

\section{Delayed self-feedback and delayed coupling}

In this section we consider the simultaneous action of delayed
coupling and delayed self-feedback.  Here we choose to apply the
self-feedback term symmetrically to both activator equations, but other feedback schemes
are also possible.

  \begin{eqnarray}\label{eq:control}
    \epsilon_1 \dot{x}_1 &=& x_1 - \frac{x_1^3}{3} - y_1+ C[x_2(t-\tau)-x_1(t)]
+K[x_1(t-\tau_K)-x_1(t)]\nonumber\\
        \dot{y}_1 &=& x_1 + a  \nonumber\\
        \epsilon_2 \dot{x}_2 &=& x_2 - \frac{x_2^3}{3} - y_2 + C[x_1(t-\tau)-x_2(t)]
+K[x_2(t-\tau_K)-x_2(t)]\nonumber\\
        \dot{y}_2 &=& x_2 + a 
    \end{eqnarray}
    
By a linear stability analysis similar to Sect. 4.1 it can be shown that the fixed point remains stable for
all values of $K$ and $\tau_K$ in case of $a>1$, as without self-feedback. Redefining
$\xi=1-a^2-C-K$, one obtains the factorized characteristic equation 
\begin{equation}
1- \xi \lambda+\epsilon \lambda^2 = \lambda K e^{-\lambda \tau_K} \pm \lambda C e^{-\lambda \tau }
\label{char_eq_K}
\end{equation}
Substituting the Hopf condition $\lambda=i \omega$ and separating into real and imaginary parts yields for the 
imaginary part
\begin{equation}
-\xi =K \cos(\omega \tau_K) \pm C \cos(\omega \tau)
\end{equation}
This equation has no solution for $a>1$ since $|\xi|=a^2-1+C+K>C+K$.

The adopted form of control allows for the synchronization of the two cells
not only for identical values of 
$\tau$ and $\tau_K$, but generates an intricate pattern of
synchronization islands or stripes in the control parameter plane (Fig.~\ref{fig:isiplane})
corresponding to single-spike in-phase and antiphase oscillations with constant interspike intervals, 
see also Fig.~\ref{fig:controlcases}(a)-(d).
Further, for adequately chosen parameter sets of coupling and self-feedback control,
we observe effects such as bursting patterns Fig.~\ref{fig:controlcases}(f) and oscillator death
Fig.~\ref{fig:controlcases}(e).
In addition to these effects, there exists a control parameter regime
in which the self-feedback has no effect on the oscillation periods (shaded yellow).

Fig. \ref{fig:isiplane} shows the control parameter plane for coupling parameters
of the uncontrolled system 
in the oscillatory regime ($C=0.5$ and $\tau=3$).  We observere three principal regimes:  
(i) Control has no effect on the oscillation period (yellow), although the form of 
the stable limit cycle is slightly altered (Fig. \ref{fig:controlcases}(a)).  
(ii) Islands of in-phase and antiphase synchronization (color coded, 
see Fig.~\ref{fig:controlcases} (b)-(d)).
(iii) Oscillator death (black) Fig.~ \ref{fig:controlcases} (e)).  

Fig.~\ref{fig:aps} shows the average phase synchronization time as a 
function of the coupling delay $\tau$ and self-feedback delay $\tau_K$
for fixed $K=0.5$. The bright straight rays at rational 
$\tau_K/\tau$ indicate long intervals during which both subsystems remain
synchronized. A particularly long average synchronization time
is found if the two delay-times are equal.

\section{Conclusion}

Our analysis has focussed on a model of two coupled neurons, which may
be viewed as a network motif for larger neural networks.  We have
shown that delayed feedback from other neurons or self-feedback from
the same neuron can crucially affect the dynamics of coupled neurons.
In case of noise-induced oscillations in instantaneously coupled
neural systems, time-delayed self-feedback can enhance or suppress
stochastic synchronization, depending upon the delay time.  This
  offers promising perspectives with respect to potential therapies of
  pathological neural synchrony as occurring, for example, in
  Parkinson's disease.  It suggests that by carefully choosing the
  delay time, feedback control applied {\it locally} to a neural
  sub-population can suppress the {\it global} synchronization of the
  neurons.

In case of delay-coupled neurons without driving noise sources, the
propagation delay of the spikes fed back from other neurons can induce
periodic oscillations for sufficiently large coupling strength and
delay times.  Bistability of a fixed point and limit cycle
oscillations occur even though the single excitable element displays
only a stable fixed point.  The two neurons oscillate with a phase lag
of $\pi$.  If self-feedback is applied additionally, for example
  by axonal reflections (Fig.~\ref{fig:scheme} (a)) in networks of electrically coupled pyramidal
  cells \cite{SCH01e}, synchronous zero-lag oscillations can be
induced in some ranges of the control parameters, while in other
regimes antiphase oscillations or oscillator death as well as more
complex bursting patterns can be generated.

\section{Acknowledgements}
This work was supported by DFG in the framework of Sfb 555. The
authors would like to thank S. Brandstetter, V. Flunkert, A. Panchuk,
and F. Schneider for fruitful discussions, and Roger Traub for
helpfull comments on gap junction coupling.


\end{document}